\documentclass[%
 reprint,
superscriptaddress,
 amsmath,amssymb,
pra,
]{revtex4-1}

\usepackage{graphicx}
\usepackage{dcolumn}
\usepackage{bm}
\usepackage{lipsum}
\usepackage[caption=false]{subfig}

\begin{document}

\preprint{Phys. Rev. A}

\title{Dielectronic resonances of LM$n$ and LN$n$ ($n$ $\geq$ 4) series in highly-charged M-shell tungsten ions}

\author{Dipti}
\email {fnu.dipti@nist.gov (Dipti)}
\affiliation{National Institute of Standards and Technology, Gaithersburg, MD, 20899, USA}

\author{A. Borovik, Jr.}
\altaffiliation[Present address: ]{I. Physikalisches Institut, Justus-Liebig-Universit\"{a}t, Giessen 35392, Germany.}
\affiliation{National Institute of Standards and Technology, Gaithersburg, MD, 20899, USA}

\author{R. Silwal}
\altaffiliation[Present address: ]{TRIUMF, 4004 Wesbrook Mall, Vancouver, BC V6T 2A3, Canada.}
\affiliation{National Institute of Standards and Technology, Gaithersburg, MD, 20899, USA}
\affiliation{Department of Physics and Astronomy, Clemson University, Clemson, SC, 29634, USA}

\author{J.M. Dreiling}
\altaffiliation{Present address: Honeywell Quantum Solutions, Broomfield, CO 80021, USA.}
\affiliation{National Institute of Standards and Technology, Gaithersburg, MD, 20899, USA}

\author{A.C. Gall}
\affiliation{National Institute of Standards and Technology, Gaithersburg, MD, 20899, USA}
\affiliation{Department of Physics and Astronomy, Clemson University, Clemson, SC, 29634, USA}

\author{E. Takacs}
\affiliation{National Institute of Standards and Technology, Gaithersburg, MD, 20899, USA}
\affiliation{Department of Physics and Astronomy, Clemson University, Clemson, SC, 29634, USA}

\author{ Yu. Ralchenko}
\email {yuri.ralchenko@nist.gov (Yu. Ralchenko)}
\affiliation{National Institute of Standards and Technology, Gaithersburg, MD, 20899, USA}

\date{\today}

\begin{abstract}
    We present spectroscopic measurements and detailed theoretical analysis of inner-shell LM$n$ and LN$n$ ($n$ $\geq$ 4) dielectronic resonances in highly-charged M-shell ions of tungsten. The x-ray emission from W$^{49+}$ through W$^{64+}$ was recorded at the electron beam ion trap (EBIT) facility at the National Institute of Standards and Technology (NIST) with a high-purity Ge detector for electron beam energies between 6.8 keV and 10.8 keV. The measured spectra clearly show the presence of strong resonance features as well as direct excitation spectral lines. The analysis of the recorded spectra with large-scale collisional-radiative (CR) modeling of the EBIT plasma allowed us to unambiguously identify numerous dielectronic resonances associated with excitations of the inner-shell $2s_{1/2}$, $2p_{1/2}$, and $2p_{3/2}$ electrons. 

\end{abstract}

\maketitle


\section{\label{sec:level1}Introduction}
Spectra of highly-charged ions (HCIs) carry the signatures of the high-temperature plasma environment and thus provide a valuable diagnostics tool. Such studies rely on the knowledge of the atomic structure of HCIs and the understanding of their interaction with other particles (electrons, photons, and ions) in the plasma. Diagnostics include the determination of plasma parameters such as electron temperature, ion temperature, electron density, ion charge state distributions, and radiation power loss. Among the various plasma parameters, the charge state distribution is one of the most important characteristics influencing the energy balance of the high-temperature plasma \cite {Beiersdorfer2012}. Radiative power loss from such plasmas, whenever significant, is strongly affected by the ion charge state distribution.  The ion charge state distribution itself depends upon the cross sections of the involved collisional processes, mainly ionization and recombination. Dielectronic recombination (DR) is one of the prevalent atomic processes affecting the ion charge state distribution and radiative power loss. It is a resonant process involving the formation of an intermediate doubly-excited autoionizing state by electron capture from the continuum, while the stabilization takes place through the emission of a photon \cite{Bauche2009}. 

Astrophysical and laboratory plasmas, such as in electron beam ion traps (EBITs) and fusion devices, are important sources of HCIs. Advanced experimental facilities along with complementing theoretical developments, motivated by numerous applications in science and technology \cite{Gillaspy2007,Beiersdorfer2015,Ralchenko}, have greatly enhanced our understanding of the physics of HCIs. An example of a pressing technological application is the study of HCIs produced in magnetically confined fusion devices \cite {Beiersdorfer2015}, e.g., tokamaks, designed for the abundant production of clean and safe energy. One of the technical challenges in achieving this goal involves the issues caused by the interaction of the hot fusion plasma with the material of the chamber walls, particularly in the divertor region \cite{Janeschitz2001}. The plasma-facing components in present day tokamaks are primarily made of tungsten (Z$_N$ = 74) due to its desirable properties, such as its high melting point and thermal conductivity, as well as low tritium retention and erosion rate. Devices of this kind include the Joint European Torus (JET) \cite {Pamela2007,Matthews2009, Hirai2007, Putterich2013}, Axially Symmetric Divertor Experiment (ASDEX) \cite{Putterich2010, Putterich2013, Neu2013},  Alcator C-Mod \cite{Greenwald2014}, and the future tokamak ITER \cite{Shimada2009,Merola2014} currently under construction in France. ITER plasma diagnostics, such as the core imaging x-ray spectrometer \cite {Beiersdorfer2010} and the vacuum ultraviolet spectrometer \cite{Seon2014}, are based upon the study of emission from tungsten impurities introduced into the fusion plasma through sputtering. A 10$^{-4}$ tungsten concentration relative to the electron density will cause unacceptable radiative power loss in the fusion plasma which can consequently prohibit the sustainable operation of the reactor \cite{Phil2011}. 

Dielectronic recombination has been studied extensively due to its direct impact on the calculation of the ion charge state distribution and radiative power losses. For example, DR  in various highly charged W$^{Z+}$ ions (Z = 18-20, 49-56, 60-72) has been studied experimentally at the heavy-ion storage ring \cite{Spruck2014, Badnell2016, Schippers2011, Badnell2012, Krantz2017} as well as at EBITs \cite{Wata2007, Biedermann2009, Tu2016, Tu2017}. Theoretical work \cite{Kwon2018a, Li2012, Li2014, Li2016, Meng2008, Wu2015, Wu2015a, Preval2019, Ballance2010, Safronova2016, Safronova2015, Peleg1998, Behar1999, Behar1999a} includes the calculations of DR rate coefficients for W$^{Z+}$ ions (Z = 1-13, 27-73) using the AUTOSTRUCTURE \cite {Badnell2011}, Hebrew University-Lawrence Livermore Atomic code (HULLAC) \cite {Bar2001}, Flexible Atomic Code (FAC) \cite {Gu2008}, Relativistic Many-Body Perturbation Theory (RMBPT) \cite{Safronova1999}, and COWAN \cite {Cowan1981} codes. The current status of theoretical and experimental work on DR data for a number of ionization stages of tungsten can be found in the recent comprehensive compilation by Kwon \textit{et al}. \cite{Kwon2018}  and references therein.  Despite the significant efforts devoted to the investigation of the DR process in various tungsten ions, the experimental and theoretical work is still insufficient to meet the data requirements for ITER diagnostics. 

Dielectronic resonances have also been studied using the EBIT facility at the National Institute of Standards and Technology (NIST). For example, LMM dielectronic resonances and radiative recombination (RR) features were identified and analyzed for Sc-like and Ti-like barium ions through measurements and theoretical calculations by McLaughlin \textit{et al}. \cite{McL1996}. LMN dielectronic resonance measurements for 3$d^n$ tungsten ions were reported through the intensity ratio of magnetic dipole lines, while detailed analysis was achieved using non-Maxwellian collisional-radiative (CR) simulations \cite{Ralchenko2013a}. In this paper, we extend our previous analysis on M-shell tungsten ions to study the inner-shell LM$n$ and LN$n$ ($n \geq$ 4) dielectronic resonances involving the experimental effort at the NIST EBIT. One of the goals is to provide benchmark data for the verification of calculations produced by different theoretical approaches. Detailed analysis of simulations of the EBIT plasma using the non-Maxwellian NOMAD code \cite{Ralchenko2001} will be presented in the following sections.

\section{Experiment}

The NIST EBIT is a unique facility for spectroscopy of HCIs \cite{Gillaspy1997}. An electron beam is produced by a Pierce-type electron gun with a Ba dispenser cathode. It is accelerated by a set of axially symmetric electrodes towards the central drift-tube region where it is then guided and compressed by a magnetic  field produced by liquid helium-cooled superconducting Helmholtz coils. A current of 147.8 A creates a magnetic field of 2.7 T, which yields a compressed electron beam with a diameter of about 35 $\mu$m and electron densities of 10$^{11}$ cm$^{-3}$ to 10$^{12}$ cm$^{-3}$. The HCIs are created and trapped in the drift-tube region, which consists of three cylindrically-shaped electrodes. The axial trapping is realized by applying a lower voltage to the middle drift tube than the two outer drift tubes, thus creating an electrostatic potential well. Ions are trapped in the radial direction by the space charge of the intense electron beam. Accessible electron energies generally range from around 200 eV up to 30 000 eV and are set by the potential difference between the cathode and the middle drift tube. After passing through the drift tube region, the electron beam is decelerated and terminated in a liquid nitrogen-cooled collector.

In the present experiment, the electron beam energy was systematically varied from 7 keV to 11 keV in steps of 50 eV, while the beam current was fixed at 150 mA. At such high current values, the space charge of the electron beam significantly influences the interaction energy of the electrons. This correction requires an additional calibration of the experimental beam energy scale. To this end, the experimental data were compared with the theoretical spectra, and the resulting correction was on the order of 200 eV for a 7 keV electron beam energy.

Tungsten ions (typically singly-charged) were injected into the drift-tube region from a Metal Vapor Vacuum Arc (MeVVA) ion source \cite{Holland2005}. It is also possible to inject gaseous elements through a ballistic neutral gas injector \cite{Fahy2007}. Due to the expected steady accumulation of impurity ions (mainly traces of barium from the cathode and xenon absorbed in the ion pumps), the trap was dumped and refilled with "fresh" tungsten ions from the MeVVA at 10 s intervals. This time scale is long compared to the fraction of a second required to create the high ion charge states of interest. It is thus reasonable to assume that the spectra accumulated for a continuous three-minute interval represents the steady-state plasma emission.

\begin{figure*}
	\centering
	\includegraphics[width=\linewidth]{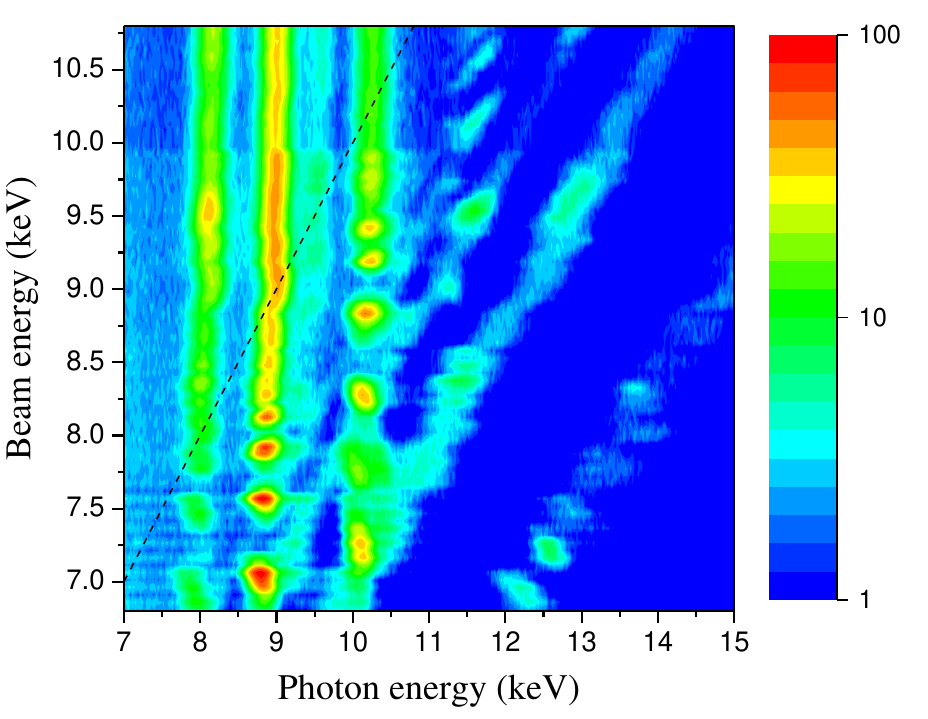}
	\caption{Measured x-ray emission spectra (in arbitrary units) using an HPGe solid state detector at electron beam energies between 6.8 keV and 10.8 keV. The dashed line separates the resonance and direct excitation features.}
	\label{fig:1}
\end{figure*}

The x-ray photons in the energy range of about 1 keV to 20 keV were detected by a high-purity germanium (HPGe) solid-state detector oriented perpendicular to the electron beam direction. The detector has a 10 mm$^2$ absorption element situated at about 20 cm from the center of the trap, and it is attached to one of the side observation ports of the EBIT. The x-ray sensor is protected by a 5 $\mu$m thick polymer window. The energy resolution of the detector is about 135 eV at 5.9 keV and changes linearly with the photon energy as an intrinsic property of solid-state x-ray detectors. The spectra were calibrated using the well known He-like lines of Ar and the direct excitation lines of tungsten ions at the nominal beam energy of 10.44 keV, which is far from any strong resonances. \par

The experimental spectra in the range of electron beam energies $E_b$ = 6.8 keV to 10.8 keV is presented in Fig. \ref{fig:1}. Although the HPGe detector collects x-ray photons well outside the presented photon energy range of $E_{ph}$ = 7 keV to 15 keV, the spectrum is zoomed in to this interval to emphasize the region of interest. Note that the signal above $E_b \approx$ 10.1 keV is weaker due to a shorter experimental collection time. While a detailed discussion of the measured spectral features will be presented in section \ref{SecAn}, one can clearly see the rich structure and converging series of resonance features (e.g., near $E_{ph} \approx$ 8.8 keV or 10.1 keV). The diagonal bands correspond to radiative recombination (free-bound) emission, and the continuous vertical bands are due to bound-bound transitions. The dashed line corresponds to $E_b = E_{ph}$, so emission below this line is solely due to dielectronic resonances which are the subject of this study.

\section{Collisional-radiative modeling}

The spectral emission recorded in EBIT experiments primarily results from interactions between the beam electrons and the trapped ions. Unlike Maxwellian plasmas, where electrons of all energies are present, the electron energy distribution function (EEDF) in an EBIT is quasi-monoenergetic. Such EEDF brings about an ionization distribution that is quite different from that in typical laboratory and astrophysical plasmas. The main difference is that the ions with ionization energy, $I$, greater than the beam energy, $E_b$, cannot be ionized (except for a very small contribution due to ionization from the lowly-populated excited states). Therefore, one can safely assume that for the range of $E_b$ = 6.8 keV to 10.8 keV, the most abundant ions of tungsten are between Mn-like W$^{49+}$ and Ne-like W$^{64+}$. \par

To accurately analyze emission from all of those ionization stages, one has to build an extensive CR model that accounts for the most important physical processes affecting atomic state populations and determines spectra for the non-Maxwellian plasma of an EBIT. In this work, we utilize the CR code NOMAD \cite{Ralchenko2001} that has been extensively used for spectroscopic diagnostics of various plasmas, e.g., EBITs, tokamaks, and laser-produced plasmas. In a general case, NOMAD calculates the rates for the prescribed EEDF and particle density using previously-generated atomic data for elementary atomic processes, solves the time-dependent first-order system of differential rate equations to deduce the state populations, and produces the synthetic spectra. In addition to the basic atomic processes describing interactions between the trapped ions and beam electrons, our CR model also takes into account the charge exchange (CX) between ions and neutral particles in the trap. 

The basic atomic data for NOMAD simulations, including energy levels, radiative decay rates, and collisional cross sections, were generated with the FAC \cite {Gu2008}. The detailed balance principle was used to obtain the cross sections for all reverse processes. The autoionization probabilities were also generated from FAC with the dielectronic capture cross sections again derived from the detailed balance. The electron-impact excitation cross sections were calculated from the oscillator strengths using the van Regemorter approximation \cite{Ragemorter1962}. This simple but computationally effective approach is justifiable here since the inner-shell resonances are produced by dielectronic capture rather than direct excitation. This set of physical processes allows us to completely account for the most important processes affecting the autoionizing state populations and the resulting x-ray spectra. The rate coefficients for all processes were obtained by integrating the calculated cross sections over the Gaussian EEDF with full width at half maximum (FWHM) of 40 eV representing the EBIT beam profile \cite{Ralchenko2013a}. The rate equations were then solved on an energy grid from 6.8 keV to 11.0 keV with steps of $\Delta E_b$ = 50 eV and at the electron density of $n_e$ = 10$^{11}$ cm$^{-3}$, and the level populations, ion charge state distributions, and x-ray spectra were subsequently generated.  \par

The starting point of any CR model is the selection of a proper representation of the atomic system in question that, on one hand, is detailed enough to describe all (or the most important) spectral features and, on the other, is tractable by the available computational resources. To analyze the sensitivity of our simulations to the model size and to its level of detail, we introduced two different models that are presented below.

\subsection{Model I}
	
The atomic states in this model were represented by relativistic configurations (RC) using the Unresolved Transition Array (UTA) mode of FAC. In the RC approach, a configuration splits into subarrays which are averaged over the fine-structure levels. For example, the configuration $1s^22s^22p^53p$ has 10 fine-structure (FS) levels due to the spin-orbit interaction, while it splits into 4 levels ($1s^22s^22\overline{p}^22p^33\overline{p}$, $1s^22s^22\overline{p}^22p^33p$,  $1s^22s^22p^42\overline{p}3\overline{p}$, and  $1s^22s^22p^42\overline{p}3p$) in the RC approach. For the representation of configurations, we use the relativistic notation throughout, where  $n\overline{l}$ describes the shell with total angular momentum, $j$ = $l$ $-$ 1/2, and  the $nl$ notation corresponds to $j$ = $l$ + 1/2. 

For the M-shell ions, the included configurations were (i) the ground configuration and excited configurations with single excitations within and from $n$ = 3 to $n$ = 4 $-$ 15, (ii) the double excitations within the M shell, and (iii) the autoionizing states produced by single excitations from the L shell ($n$ = 2) to $n$ = 3 $-$ 15 and the double excitations from the L and M shells to the 4$lnl'$ ($n$ = 4 $-$ 8) and 5$l$5$l'$. The model comprises electric dipole (E1) transitions among all the configurations and also includes magnetic dipole (M1), electric quadrupole (E2), and magnetic quadrupole (M2) transitions between configurations involving single and double excitations within the M shell. Model I includes a total of approximately 0.11 million states and 1.8 million radiative transitions. 

In order to keep computations at a manageable level, we implemented a ``sliding window" approach where the range of ions included in the calculation shifts with the electron beam energy. For example, to describe the emission from Cr-like and V-like ions, only ion charge states ranging from Mn-like W$^{49+}$ to Ti-like W$^{52+}$ were included in spectrum calculations. Even with this restriction, the CR model is quite large as it includes approximately 39 000 states and 0.6 million radiative transitions. Although a sliding window of only four ion charge states is rather narrow, this approach still allows accurate calculation of ionization distributions and the corresponding spectra. In low-density plasmas, the ionization balance is established through ionization and recombination processes between the adjacent ion stages; i.e., 
  \begin{equation}
  \frac{N_{Z+1}}{N_Z} = \frac{R_I}{R_{RR} + R_{DR} + R_{CX}}.   
  \end{equation}
Here, $N_Z$ represents the ion population, and $R_I$, $R_{RR}$, $R_{DR}$, and $R_{CX}$ are rates of ionization, radiative recombination, dielectronic recombination, and charge-exchange, respectively. Therefore, the {\it relative} intensities of the calculated spectra for the two middle ionization stages (for instance,  between the Cr-like and V-like ions in the example above) were adequately determined. When the sliding window for calculations at a fixed $E_b$ is shifted to the next group of ions, the relative line intensity ratios for the next pair of ions is again correctly calculated. At the end of the simulations, the total ion populations, $N_Z$, as well as the state populations were renormalized according to $\Sigma_{Z}{N_Z}=1$, and thus this procedure resulted in a consistent determination of the synthetic x-ray spectra. 

\begin{figure*}
		\centering
		\includegraphics[width=0.8\linewidth]{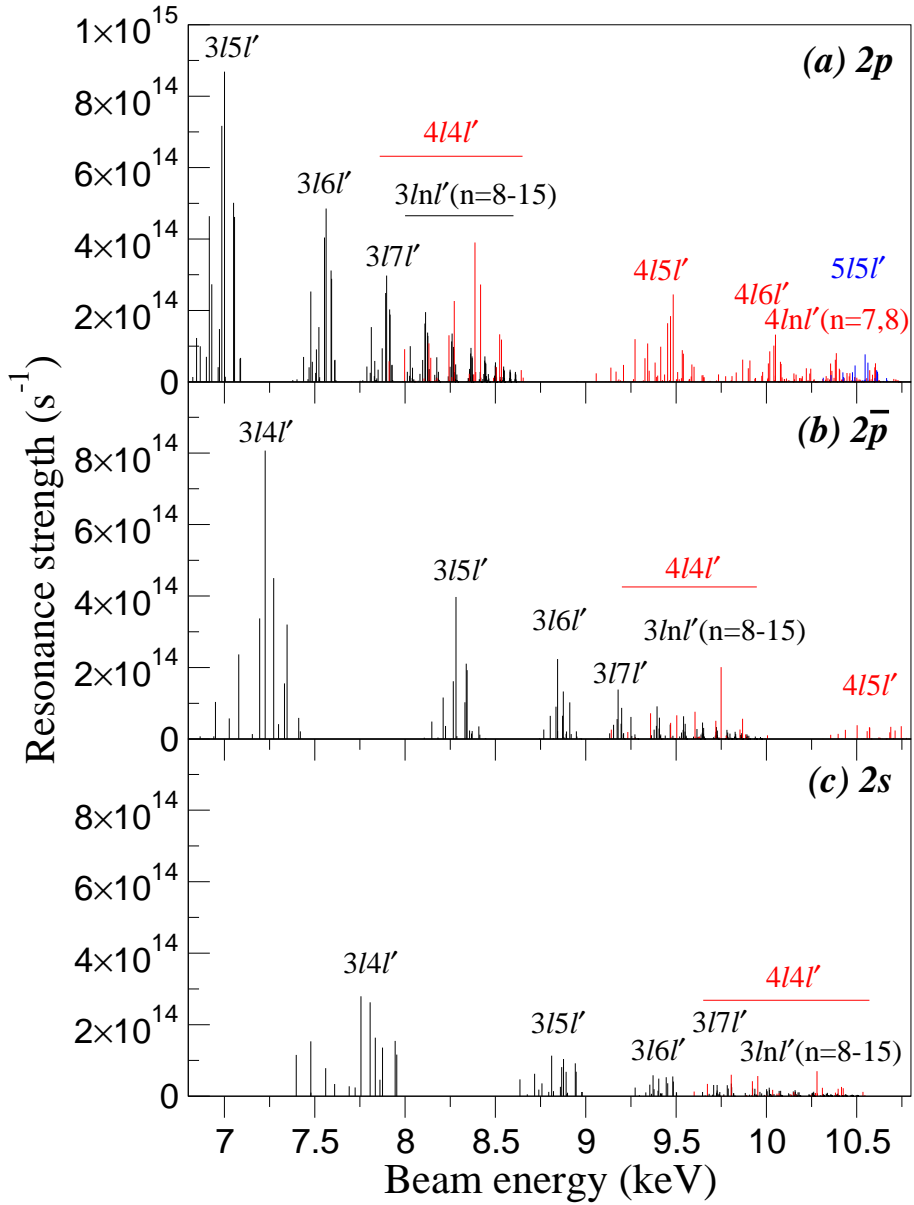}
		\caption{Resonance strengths (in s$^{-1}$) for the dielectronic capture from the ground state of Ar-like W$^{56+}$ forming the doubly-excited states of K-like W$^{55+}$ ions. Labels correspond to the excitation of (a) 2$p$, (b) 2$\overline{p}$, and (c) 2$s$ electron into the $nl$ ($n$ = 3 $-$ 5)  shell with the simultaneous capture of an electron into the  other shell $n'$. Black, red, and blue lines represent the different resonances corresponding to L-shell excitation to M (LM$n'$), N (LN$n'$), and O (LO$n'$) shell, respectively.}
		\label{fig:2}
	\end{figure*}

 Figure \ref{fig:2} presents an example of the detailed resonance strengths for electron capture from the ground state (3$p^6$) of Ar-like W$^{56+}$ forming the doubly-excited states of K-like ions. LM$n$ ($n$ $\geq$ 4), LN$n$ ($n$ $\geq$ 4), and LO$n$ ($n$ = 5) resonances are produced when the L electron is excited to the M, N, and O shell, respectively (leaving a hole in a 2$p$, 2$\overline{p}$, or 2$s$ orbital), and simultaneously the free electron is captured into an atomic shell with a principal quantum number $n$. In Fig. \ref{fig:2}, resonance strengths corresponding to excitation from the 2$p$, 2$\overline{p}$, and 2$s$ orbital are presented as a function of the electron beam energy. Several observations can be made from Fig. \ref{fig:2}. 1) The resonance strengths show the expected decrease with increasing $n$ as a function of beam energy. 2) L$_p$M$n$ resonance strengths are at least a factor of two larger than L$_{\overline{p}}$M$n$ resonances having a 2${\overline{p}}$ hole in the $n$ = 2 shell. 3) Higher dielectronic resonances ($n$ $\geq$ 8) are immersed into one broad structure. 4) Resonances from the same configuration are spread over energies corresponding to states with different $lj$ quantum numbers.  

\subsection{Model II for K-like W$^{55+}$}

To study the DR process at the most detailed level, we also developed a model for a particular charge state where the atomic structure is represented by fine-structure levels rather than relativistic configurations.  The model includes non-autoionizing as well as autoionizing states of the K-like W$^{55+}$ and non-autoionizing states of the Ar-like W$^{56+}$ ions. K-like ions only have one valence electron in their ground configuration of 2\emph{s}$^2$2$\overline{p}$$^2$2\emph{p}$^4$3\emph{s}$^2$3$\overline{p}$$^2$3\emph{p}$^4$3$\overline{d}$. Table \ref{table:1} presents the configurations included for the K-like W$^{55+}$ ion, which resulted in approximately 17 000 FS levels. We also performed simulations for the K-like tungsten ions in the RC approach for the same set of configurations, this time resulting in about an order of magnitude smaller number of levels as compared to FS levels. Approximately 40 million radiative transitions between the different fine-structure levels of K-like ion were taken into account. This is nearly 22 times more than the total number of radiative transitions for all ions in Model I.

\begin{table}[t]
		\caption{Configurations for K-like W$^{55+}$ ion included in Model I. The principal quantum number is represented by $n$ = 4 $-$ 9 and $n'$ = 4 $-$ 8. Notations 2$l^k$ correspond to all possible permutations of $k$ electrons in the L shell.}
		\centering
		\begin{tabular}{lcl}
			\hline\hline
			M-shell excitations &  & L-shell excitations  \\ [0.5ex]
			\hline\hline
			2$l^8$3\emph{s}$^2$3\emph{p}$^6$3\emph{d} &  & 2\emph{l}$^7$3\emph{s}$^2$3\emph{p}$^6$3\emph{d}$^2$ \\
			2$l^8$3\emph{s}$^2$3\emph{p}$^5$3\emph{d}$^2$ &  & 2\emph{l}$^7$3\emph{s}$^2$3\emph{p}$^6$3\emph{d}n\emph{l} \\
			2$l^8$3\emph{s}3\emph{p}$^6$3\emph{d}$^2$ &  & 2\emph{l}$^7$3\emph{s}$^2$3\emph{p}$^6$4\emph{l}n$'$\emph{l}$'$  \\
			2$l^8$3\emph{s}$^2$3\emph{p}$^4$3\emph{d}$^3$ &  & 2\emph{l}$^7$3\emph{s}$^2$3\emph{p}$^6$5\emph{l}5\emph{l}$'$\\
			2$l^8$3\emph{s}3\emph{p}$^5$3\emph{d}$^3$ & &\\
			2$l^8$3\emph{p}$^6$3\emph{d}$^3$ &&\\
			2$l^8$3\emph{s}$^2$3\emph{p}$^6$n\emph{l} &&\\
			2$l^8$3\emph{s}$^2$3\emph{p}$^5$3\emph{d}n\emph{l}  &&\\
			2$l^8$3\emph{s}3\emph{p}$^6$3\emph{d}n\emph{l} &&\\ [1ex]
						\hline
		\end{tabular}
		\label{table:1}
	\end{table}

\section{Analysis and results\label{SecAn}}
\subsection{Relativistic configurations vs. fine-structure}

\begin{figure*}
\begin{minipage}[t]{0.58\linewidth}
\includegraphics[width=\linewidth, height=8cm]{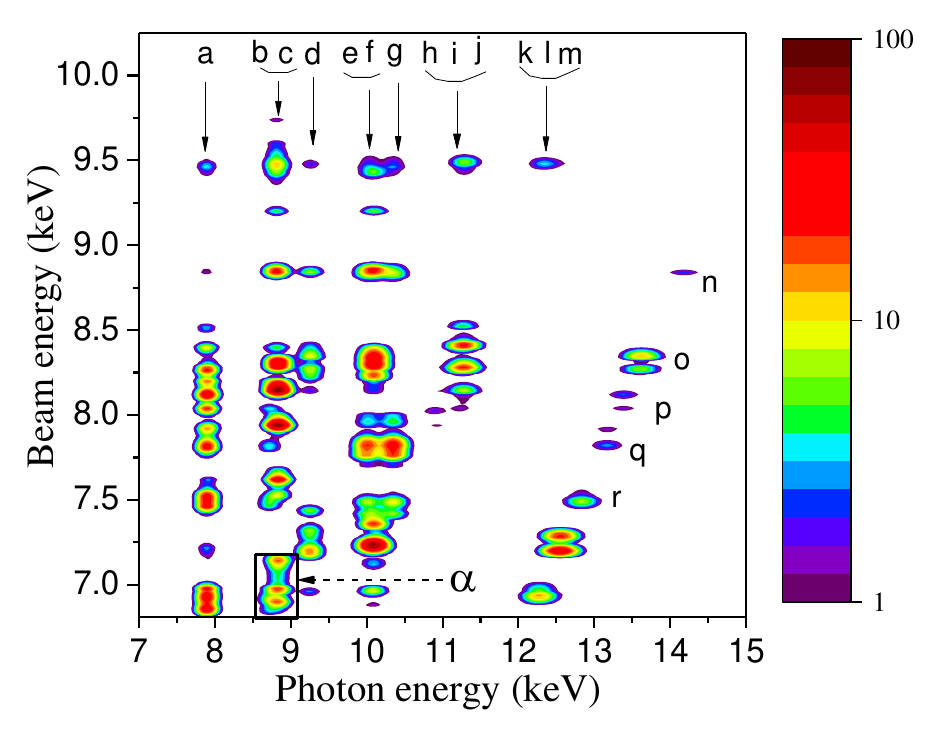}
\caption{Theoretical x-ray emission spectra for K-like W$^{55+}$ ion using fine-structure levels. The area labelled by '$\alpha$' shows the emission from the decay of the 2$p^5$3$d$5$l$ doubly-excited states. The letters (a to r) in the figure represent the following x-ray transitions: \textbf{a}: 2$p$ $-$ 3$s$;  \textbf{b}: 2$p$ $-$ 3$\overline{d}$; \textbf{c}: 2$p$ $-$ 3$d$; \textbf{d}: 2$\overline{p}$ $-$ 3$s$; \textbf{e}: 2$s$ $-$ 3$\overline{p}$; \textbf{f}: 2$\overline{p}$ $-$ 3$\overline{d}$; \textbf{g}: 2$s$ $-$ 3$p$; \textbf{h}: 2$p$ $-$ 4$s$;  \textbf{i}: 2$p$ $-$ 4$\overline{d}$; \textbf{j}: 2$p$ $-$ 4$d$; \textbf{k}: 2$p$ $-$ 5$s$;  \textbf{l}: 2$p$ $-$ 5$\overline{d}$; \textbf{m}: 2$p$ $-$ 5$d$; \textbf{n}: 2$\overline{p}$ $-$ 6$\overline{d}$;  \textbf{o}: 2$\overline{p}$ $-$ 5$\overline{d}$; \textbf{p}: 2$p$ $-$ 8$d$; \textbf{q}: 2$p$ $-$ 7$d$; and \textbf{r}: 2$p$ $-$ 6$d$.}
\label{fig:3}
\end{minipage}\hfill%
\begin{minipage}[t]{0.38\linewidth}
\includegraphics[width=\linewidth, height=9cm]{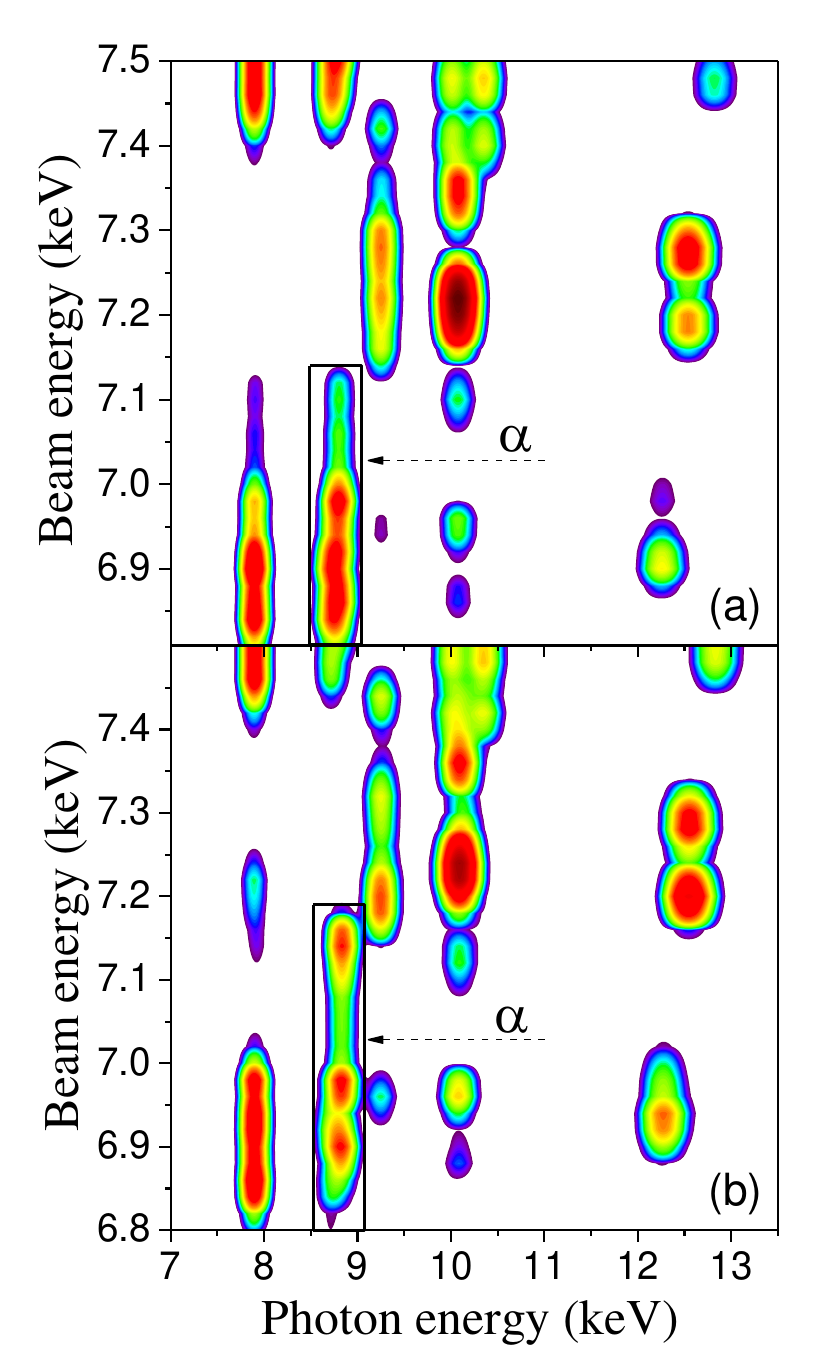}
\caption{Higher resolution comparison of the region '$\alpha$' (Fig. \ref{fig:3}) for the simulations performed using (a) relativistic configurations and (b) fine-structure levels. X-ray emissions in this region are due to the 2$p $ $-$ 3$\overline{d}$ and 2$p$ $-$ 3$d$ transitions from the radiative decay of 2$p^5$3$d$5$l$ doubly-excited states. The intensity scale is same as used in Fig. \ref{fig:3}.}
\label{fig:4}
\end{minipage}%
\end{figure*}

\begin{figure}
	\centering
	\includegraphics[width=\linewidth]{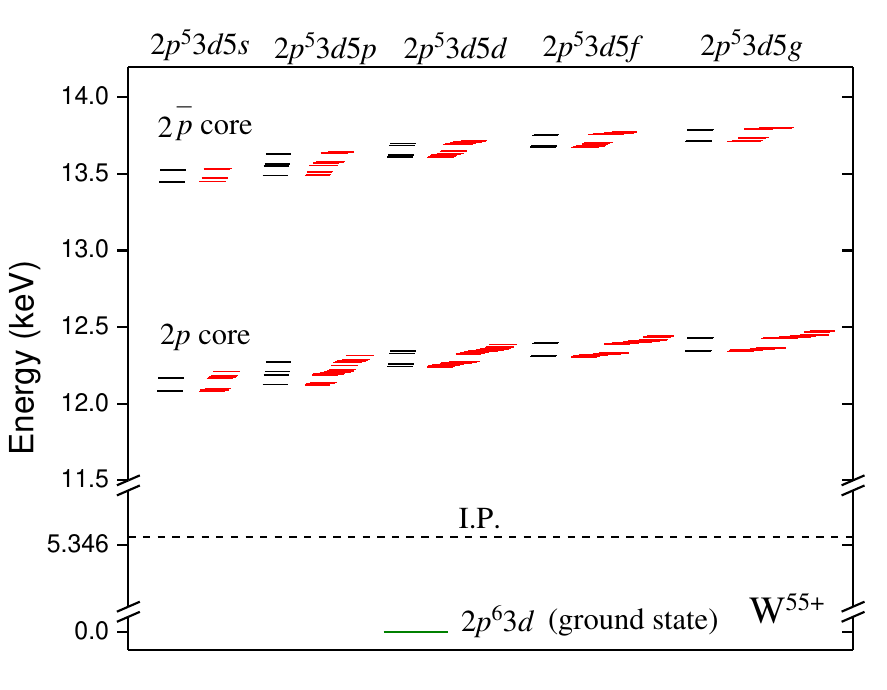}
	\caption{Energy level diagram of the K-like W$^{55+}$ ion for the configuration 2$p^5$3$d$5$l$ relative to its ground state in relativistic configurations (black lines) and fine-structure (red lines) mode FAC calculations.}
	\label{fig:5}
\end{figure}

The relativistic configurations and fine-structure calculations  predicted similar emission features for K-like tungsten ions; therefore, only the theoretical x-ray spectra obtained using the fine-structure levels are shown in Fig. \ref{fig:3}. X-ray lines due to the different radiative stabilizing channels for LM$n$ and LN$n$ autoionizing states are labelled in the same figure.

In Fig. \ref{fig:3}, an area '$\alpha$' shows the emission following the decay of the 2$p^5$3$d$5$l$ doubly-excited states. Figure \ref{fig:4} shows this area with higher resolution for comparison of the RC and FS simulations.  Slight differences in the resonance energies and the intensities were observed between the two spectra. To understand the differences in the spectra obtained from the RC and FS calculations, let us compare the energy level diagram of the 2$p^5$3$d$5$l$ configuration in the two approaches (Fig. \ref{fig:5}). Due to the spin-orbit interaction, this configuration splits into two groups with a 2$p$ and a 2$\overline{p}$ hole in the $n$ = 2 shell. These levels are separated by an energy of approximately 1.4 keV in both approaches. The magnitude of the spin-orbit interaction is much smaller for the coupling of the 2$p$ or 2$\overline{p}$ states of the 2$p^5$ ion core with the 3$d$ and 3$\overline{d}$ orbitals (separated by only a few tens of eVs) and even smaller for outer electrons. Energies of the levels generated from the coupling of the 2$p$ core with 3$d$5$l$ in the RC approach are lower than the corresponding energies of FS levels by at most 45 eV. This difference in the energy levels of the RC and FS calculations is reflected in the resonance energies in the marked area '$\alpha$' of Fig. \ref{fig:4}. The intensity differences between the two models (Fig. \ref{fig:4}) can be attributed to the fact that the rates are averaged over statistical weights in the relativistic configurations. Photon energies convolved with experimental Gaussian shapes (FWHM $\approx$ 135 eV at 5.9 keV) are similar in the two calculations.  \par 

Overall, the differences between the two calculations are not very significant. This leads us to the conclusion that the relativistic configuration approach is sufficient to describe most of the observed features with reasonable accuracy and, at the same time, keeps the computational resources at a manageable level.

\subsection{Comparison of the experimental spectra with theory}

\begin{figure*}
	\centering
	\includegraphics[width=0.8\linewidth,height=\linewidth]{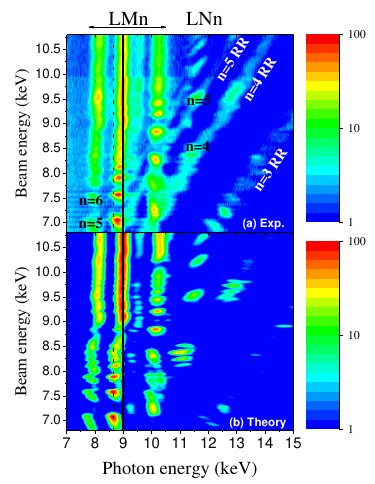}
	\caption{Comparison of the (a) experimental spectra and (b) theoretical x-ray emission spectra obtained from Model I. X-ray emission due to excitation and DR appear along vertical bands. Strongest dielectronic resonances LM$n$ and LN$n$ are labelled. The diagonal bands correspond to RR emission. Bright spots along  the diagonal bands are DR features. Solid and dashed lines, centered on the theoretical direct excitation and DR photon energies, respectively, show the differences in the experimental and theoretical energies.}
	\label{fig:6}
\end{figure*}

Figure \ref{fig:6} compares the measured spectra to the results of RC simulations. As mentioned earlier, the vertical bands in the experimental spectra correspond to the DR and direct excitation x-ray transitions. The diagonal bands are due to RR transitions to the $n$ = 3$-$5 shells. For instance, the photon emission near 12.3 keV at the beam energy of 7 keV results from recombination of the free electron into the n = 3 shell of the K-like ion which has a binding energy of about 5.3 keV. The RR contributions can be described reasonably well with the existing theoretical methods. Therefore, below we will focus on the dielectronic resonances only and completely omit RR contributions from the theoretical spectra. \par

 The strongest DR lines in the measured spectrum originate from the LM$n$ ($n$ $\geq$ 4) autoionizing states produced by the excitation of the 2$l$ electron into the 3$l'$ shell (2$l$ $-$ 3$l'$) with the simultaneous capture of a continuum electron into the $n$ shell. The vertical bands at photon energies of about 7.9 keV and 8.8 keV are due to the strongest stabilizing E1 transitions 2$p$ $-$ 3$s$ and a blend of 2$p$ $-$ 3$\overline{d}$ and 2$p$ $-$ 3$d$ transitions,  respectively. The other radiative stabilizing channels of the LM$n$ levels involve the 2$\overline{p}$ and 2$s$ subshells and give characteristic emission near 9.2 keV (2$\overline{p}$ $-$ 3$s$) and 10.2 keV (a blend of 2$s$ $-$ 3$\overline{p}$, 2$\overline{p}$ $-$ 3$\overline{d}$, and 2$s$ $-$ 3$p$ transitions). Energies of the x-ray photons emitted during the DR are close to the resonance transition energies of the parent ion but are slightly shifted by the presence of the captured spectator electron into different $nl$ atomic shells. \par

Characteristic emissions that produce the strong resonance lines near 11.4 keV are the result of 2$l$ $-$ 4$l'$ transitions from LN$n$ ($n$ $\geq$ 4) autoionizing states. Alternatively, the stabilization of these doubly-excited states could also lead to the occupation of states still above the ionization energy of the recombined ion. These excited states are then further susceptible to other secondary stabilization transitions.  As an example, the 2$p^5$3$s^2$3$p^5$4$d$4$f$ (LNN) Ar-like level is produced by inner-shell dielectronic capture involving the 2$p^6$3$s^2$3$p^5$ ground state of a Cl-like ion. This doubly-exited state mainly decays into two levels, the 2$p^5$3$s^2$3$p^5$3$d$4$d$ (LMN) and the 2$p^6$3$s^2$3$p^5$4$f$ via 3$d$ $-$ 4$f$ and 2$p$ $-$ 4$d$ E1 radiative transitions, respectively.  The LMN doubly-excited state then decays into the 2$p^6$3$s^2$3$p^5$3$d$ level giving rise to a 2$p$ $-$ 4$d$ transition ($\Delta$E = 11.29 keV) or into the 2$p^6$3$s^2$3$p^5$4$d$ level resulting in a 2$p$ $-$ 3$d$ ($\Delta$E = 8.74 keV) transition. Both of these transitions are observed in the spectra.
\par

The recombined LM$n$ and LN$n$ doubly-excited states can also stabilize through the radiative de-excitation of the outer electron $nl$ to lower $n'l'$ states. The resonances observed across the $n$ = 3 and $n$ = 4 RR bands correspond precisely to this type of decay. For instance, the experimental spectra show decays of 2$p$ $-$ $ns$ ($n$ = 5-8), 2$p$ $-$ $n\overline{d}$ ($n$ = 5-10), 2$p$ $-$ $nd$ ($n$ = 5-9), 2$\overline{p}$ $-$ 5$s$, 2$\overline{p}$ $-$ $n\overline{d}$ ($n$ = 5-14), 2$s$ $-$ 5$\overline{p}$, and 2$s$ $-$ 5$p$ for x-ray energies $\gtrsim$ 12 keV.\par

It is evident that theoretical simulations successfully predict most of the observed spectral features in terms of the line positions and relative intensities. However, slight differences between the measured and simulated spectra were observed along the vertical bands. The photon energies of the direct excitation lines (solid line centered on the theoretical excitation energies in Fig. \ref{fig:6}) and DR features (dashed line centered on the theoretical DR x-rays) are not the same experimentally and theoretically. This may be due to the unknown effect of charge exchange on the ionization balance that slightly modifies the distribution of the most abundant ions and, consequently, results in somewhat shifted positions of the strongest dielectronic resonances.

Due to the unavoidable presence of neutrals in the trap, the electron (charge) exchange between highly-charged ions and neutrals always affects the ionization balance of the EBIT plasma. The CX rate between ions of charge Z and Z+1 can be approximated as
\begin{equation}
    R_{CX} = N_0 \cdot \sigma^{Z}_{CX} \cdot v_r,
\end{equation}
where $N_0$ is the density of neutral particles, $\sigma^{Z}_{CX}$ is the CX cross section from Z+1 into Z, and $v_r$ is the relative velocity between neutrals and ions. Due to the lack of CX calculations for tungsten ions with Z $\approx$ 50-60, the only practical approach  for determining the CX cross sections is to make use of the Classical Trajectory Monte Carlo recommendations of $\sigma^{Z}_{CX}$ = $Z \cdot 10^{-15}$ cm$^2$ \cite{Otranto2006}. This leaves the product $\rho_{CX} = N_0 \cdot v_r$ as the only unknown parameter that can, in principle, be derived from fitting the experimental data. We followed this strategy in our previous papers  \cite{Ralchenko2008, Ralchenko2013a}, where spectral lines from different ionization stages were well-separated, thus allowing reliable fits to the measured ionization balance and determination of $\rho_{CX}$. It was found that for typical measurements with high-Z metals, $\rho_{CX} = (1-3) \times 10^{12}$ cm$^{-2}$s$^{-1}$. In the present experiment, the spectral features from different ions strongly overlap; thus, it is not possible to derive $\rho_{CX}$ directly.  Our simulations therefore used the value of $\rho_{CX}$ = 2 $\times 10^{12}$ cm$^{-2}$s$^{-1}$ at all beam energies. It should be mentioned that the experimental conditions are not same at each beam energy and/or current, and thus the CX rates can vary. The small differences in the experimental and theoretical spectra (Fig. \ref{fig:6}) at certain beam energies may be attributed in part to the deviation of the actual CX rates from the average value in the model.

\section{Conclusions\label{SecCon}}

In this paper, we presented a detailed experimental and theoretical study of the inner-shell dielectronic resonances in the M-shell ions of tungsten.  The x-ray spectra measured on the NIST Electron Beam Ion Trap for electron beam energies between 6.8 keV and 10.8 keV revealed series of LM$n$ and LN$n$ resonances stabilizing via the $2l-3l'$ and $2l-4l'$ as well as $2l-nl'$ radiative transitions. The emission features were identified with the help of a large-scale collisional-radiative model that included sixteen ionization stages, more than 100,000 atomic states, and about 2 million radiative and collisional transitions. This comprehensive analysis generally reproduced the observed resonances and direct excitation features. Slight differences in the observed and simulated spectra may be attributed to changes in the calculated ion charge state distribution due to the small unknown contribution from the charge exchange process which is unavoidable in EBITs. 

The presented x-ray spectra were recorded with a high-purity Ge detector that can provide a rather limited energy resolution on the order of 140 eV. This is clearly insufficient to distinguish either fine-structure resonance features or CX effects. While it would be difficult to fully explore these spectroscopic signatures with high-resolution crystal spectrometers due to their narrow spectral ranges, the recent developments in multi-pixel microcalorimeters that offer both extensive energy coverage and very good energy resolution on the order of only a few eV (e.g., \cite{Szy2019}) give great hope that it will soon be  possible to reach a much better understanding of inner-shell DR features in highly-charged high-Z ions. These measurements can help facilitate more reliable modeling and prediction of ionization balance and power losses in high-Z plasmas such as those in tokamaks, laser-produced plasmas, and astrophysics.

\section*{Acknowledgements}
This work was partially funded by the NIST Grant Award Numbers 70NANB16H204, 70NANB18H284, and 70NANB19H024 of the Measurement Science and Engineering (MSE) Research Grant Programs, and by the National Science Foundation Award Number 1806494. AB acknowledges the support of the US Department of Commerce via the Guest Researcher Program. JMD acknowledges funding from a National Research Council Postdoctoral Fellowship.

%


\end{document}